\documentstyle[11pt,fullpage,doublespace,epsf]{article}
 \DeclareMathSizes{11}{19}{13}{9}   % For size 11 text

\makeatletter
\@addtoreset{equation}{section}
\makeatother

\begin{document}
   
\title{Universe Decay, Inflation and the Large Eigenvalue of the Cosmological Constant Seesaw}
\author{Michael McGuigan\\Brookhaven National Laboratory\\Upton NY 11973\\mcguigan@bnl.gov}

\date{}
\maketitle

\begin{abstract}We discuss implications of the large eigenvalue of the cosmological constant seesaw mechanism extending hep-th/0602112 and hep-th/0604108. While the previous papers focused on the small eigenvalue as a cosmological constant associated with the accelerating Universe, here we draw attention to the physical implications of the large eigenvalue. In particular we find that the large eigenvalue can give rise to a period of inflation terminated by Universe decay. The mechanism involves quantum tunneling and  mixing and introduces parameters $\Gamma$, the decay constant, and $\theta$, the mixing angle. We  discuss the cosmological constant seesaw mechanism in the context of various models of current interest including chain inflation, inflatonless inflation, string theory, Universe entanglement and different approaches to the hierarchy problem. 
\end{abstract}

\section{Introduction}

The cosmological constant is an important physical quantity because a proper understanding of this modification to the traditional Einstein-Hilbert action requires a nonperturbative formulation of quantum gravity (see for example the lucid discussion of \cite{Witten:2000zk}). To take a quote from Witten out of context \cite{Witten:2000zk}:

\begin{quotation}
"The problem of the vacuum energy or cosmological constant - why it is zero or extremely small by particle physics standards- really only arises in the presence of gravity, since without gravity, we don't care about the energy of the vacuum. Moreover it is mainly a question about quantum gravity, since classically it would be more or less natural to just decide - as Einstein did- that we do not like the cosmological constant and set it to zero." (E. Witten (2000))
\end{quotation}
Thus progress is understanding the cosmological constant is crucial if one wants to move toward the ultimate goal of quantizing gravity. Also increasingly precise measurements of dark energy are consistent with the existence of a nonzero positive cosmological constant. If they turn out to be correct it will be the first experimental measurement of any kind that requires a quantum gravity explanation. 

There are many approaches to the cosmological constant problem \cite{Weinberg:2000yb}\cite{Carroll:2000fy}\cite{Polchinski:2006gy}. For example anthropic approaches set a anthropic window on it's value \cite{Weinberg:2005fh}\cite{Weinberg:1987dv}\cite{Banks:1984cw}.  In this paper we will discuss a method of relating the value of the cosmological constant to the phenomenon of Universe decay. In doing so it will become clear that quantum gravity is important in such transitions as well as in the cosmological constant problem.

Universe decay is the transition from one Universe to another and involves a parameter $\Gamma$ which parametrizes an exponential suppression of the transition. The discussions of the phenomena of Universe decay occurred in quantum gravity descriptions of topology change (see for example \cite{McGuigan:1989yb}). Recently Universe decay has been shown to give rise to a type of inflationary model without an inflaton field whose quantitative predictions are similar to a slow-rollover inflationary model \cite{Watson:2006px}. In this paper we will discuss how to incorporate the parameter $\Gamma$ into the cosmological constant seesaw mechanism of \cite{McGuigan:2006hs} and \cite{McGuigan:2006iq}. We will then discuss the relation to other approaches such as chain inflation \cite{Freese:2006fk}, tunneling in string theory and Universe entanglement.

The cosmological constant seesaw relation is an inverse relation of the form:
\[
\lambda  = c\frac{{ m_{EW}^8 }}{{m_{P}^4 }}
\]
which illustrates a multiplicative suppression of the Planck scale cosmological constant. This is different from  additive \cite{Duff:1980qv}\cite{Jackiw:1986pg}\cite{Henneaux:1984ji}\cite{Brown:1987dd} \cite{Duncan:1989ug}\cite{Duncan:1990fr}\cite{Bousso:2000xa} and exponential suppression \cite{Svrcek:2006hf} approaches to the cosmological constant. Here\\ $m_{EW}=877GeV=2.995G_F^{-1/2}$ is related to  the   electroweak symmetry breaking scale, 
$m_{P}  = (8\pi G)^{ - 1/2}  = 2.4353 \times 10^{18} GeV$ is the reduced Planck mass and $c$ is a numerical constant. The relation is consistent with the observed value of:
\[
\lambda  = (.316meV)^4
\]

The cosmological constant seesaw relation occurs is several different contexts. The approaches include  quintessance \cite{Arkani-Hamed:2000tc}\cite{Chacko:2004ky}, gravitino seesaw relations via superhiggs interactions \cite{Barbieri:1981gn}\cite{Fayet:1979qi}\cite{Ibe:2005xc}\cite{Fayet:1986zc}, Casimir contributions to the vacuum energy \cite{Chen:2006nu}, vacuum energy expansion in supersymmetry breaking \cite{Chalmers:2005wy}, six dimensional supergravity \cite{Aghababaie:2003wz}\cite{Burgess:2004ib} and  cosmological supersymmetry breaking \cite{Banks:2000fe}. In this paper we follow the approach of Motl  \cite{Motl} where the cosmological constant seesaw is treated using a $2\times2$ matrix. This method  is less dependent on details of an underlying supersymmetry than some of the other approaches. 

Further motivation for the cosmological constant seesaw can be found in traditional particle physics relations that relate low energy to high energy mass scales through inverse multiplicative relationships. One primary example of such a relationship is neutrino mass seesaw relation \cite{Ramond:1998hs}\cite{Langacker:2005pf}:
\begin{equation}
m_\nu = c_\nu \frac{{m_{EW}^2 }}{{m_{GUT} }}
\end{equation}
Here the large scale in the denominator is approximately $m_{GUT} = 10^{15}GeV$ and yields neutrino masses in the $eV$ range. Another example is the axion mass relationship \cite{Wilczek:1977pj}\cite{Svrcek:2006yi}:
\begin{equation}
m_a  = c_a\frac{{ F_\pi  m_\pi  }}{{F_a /r}}
\end{equation}
In the above $F_a$ is a decay constant of the axion and $r$ is a parameter used to normalize it's kinetic term \cite{Svrcek:2006yi}. Here the large mass scale is taken to be $F_a /r = 1.1 \times 10^{16} GeV$ and yields $m_a  = .00054\mu eV$. 

If the cosmological seesaw relation is realized in nature one has a similar relation to (1.1) and (1.2) in that there a multiplicative suppression of the quantity with a large value in the denominator. Unlike the other two relations however the derivation of the cosmological constant seesaw relation from a fundamental point of view has proved very difficult. It is clear that the relation relates high energy with low energy physics. Because of this low energy/high energy connection short distance physics (quantum gravity) plays a major role in a proper understanding of the cosmological constant seesaw relation. The extreme low value of the constant will not be a random occurrence like in a anthropic approach.  

For the cosmological constant seesaw the low value of the cosmological constant is a fundamental feature of  short or long distance quantum gravity. This approach avoids the objection that low energy quantities like the cosmological constant cannot possibly probe high energies because the dynamics at extreme high energies are insensitive to their values. Rather when one measures low energy quantities such as the neutrino mass, axion mass or cosmological constant one is opening a window into the highest energy scales. 

It is important to stress however that despite the similarities to (1.1) and (1.2) the cosmological constant is not a mass term of the traditional form found in particle physics. Indeed $\lambda$ can be present in the Einstein-Hilbert Lagrangian and the graviton or gravitino can still be massless. Instead what happens is that in the presence of $\lambda$ the canonical structure of the theory is altered. When one quantizes the theory this modification of the canonical structure manifests itself as a Klein-Gordon type equation of the gravitation field called the Wheeler-DeWitt (WDW) equation. For certain gauge choices or clock variables the $\lambda$ term will look like a $(mass)^2$ term in this Klein-Gordon type equation. However there are other clock choices in which it does not. For theories with closed spatial topologies considered in this paper there is no unique notion of time or clock variable. In this respect traditional quantum mechanics loses out to gravity. For example a traditional Schrodinger equation requires that a clock can be defined globally.

We expect however that Universe decay transitions can be described and quantified in any clock variable without any measurable differences. Indeed the WDW equation can be interpreted as the invariance of the wave function of the Universe under the generator of time reparametrization. A stronger treatment than the WDW equation is to use a precise definition of the wave function of the Universe using string theory which is a UV complete theory \cite{Ooguri:2005vr}\cite{Gukov:2005iy}. For technical reasons this is currently limited to unbroken supersymmetry \cite{Ooguri:2005vr}\cite{Gukov:2005iy} or to two and three dimensional nonsupersymmetric models \cite{McGuigan:2006iq}.

In \cite{Strominger:2001gp} Strominger outlined the following  cosmological paradigm: Large DeSitter expansion in the early Universe, radiation and matter dominated expansion in the medium-time Universe, and small DeSitter expansion at late times.  We will try to implement the paradigm by taking  simple model of Universe decay and mixing involving  the $2\times 2$ matrix representation. In this paper we are mainly interested in positive $\lambda$ because we want to model an early inflationary Universe and a late-time accelerating Universe  ( The case of negative $\lambda$ is very different and is perhaps best described using the $AdS/CFT$ correspondence). For positive cosmological constant we set  $\Lambda = M^2$ where:
\[
M = \left( {\begin{array}{*{20}c}
   0 & {\mu _{12} }  \\
   {\mu _{12} } & {\mu _{22} }  \\
\end{array}} \right) = \left( {\begin{array}{*{20}c}
   0 & {c_{12} m_{EW}^2 }  \\
   {c_{12} m_{EW}^2 } & {c_{22} m_P^2 }  \\
\end{array}} \right)
\]
The constants $c_{12}$ and $c_{22}$ are parameters defined so that one obtains  a reasonable cosmological constant when they are of order one. The physical values of the cosmological constant are determined by the eigenvalues of the cosmological constant matrix and are denoted by $\lambda_{I,II}$. These are determined from the eigenvalues of the $M$ matrix $\mu_{I,II}$ through:
\[
\lambda _{I,II}  = \mu _{I,II}^2
\]
The eigenvalues of the $M$ matrix are given by:
\[
\begin{array}{l}
 \mu _I  = \frac{1}{2}(\mu _{22}  - \sqrt {4\mu _{12}^2  + \mu _{22}^2 } ) \approx  - \frac{{\mu _{12}^2 }}{{\mu _{22} }} \approx  - \frac{{\mu _{12}^2 }}{{\sqrt {\Lambda _{22} } }} \approx  - \frac{{c_{12}^2 }}{{c_{22} }}\frac{{m_{EW}^4 }}{{m_P^2 }} \\
 \mu _{II}  = \frac{1}{2}(\mu _{22}  + \sqrt {4\mu _{12}^2  + \mu _{22}^2 } ) \approx \sqrt {\Lambda _{22} }  \approx c_{22} m_P^2  \\
 \end{array}
\]
The physical values of the cosmological constant matrix are therefore approximately:
\[
\begin{array}{l}
 \lambda _I  = \mu _I^2  \approx \frac{{c_{12}^4 }}{{c_{22}^2 }}\frac{{m_{EW}^8 }}{{m_P^4 }} \\
 \lambda _{II}  = \mu _{II}^2  \approx c_{22}^2 m_P^4  \\
 \end{array}
\]
 We shall see in the next section that the solutions to the WDW equation will be dominated by the large eigenvalue at small $a$, where $a$ is the radius of the Universe. This contribution will be suppressed through a decay factor $\Gamma$ for large $a$ where the solution is driven by the small eigenvalue, so effectively our description of Universe decay describes a transition from large to small $\lambda$.

This paper is organized as follows. In section 2 we describe Universe decay in the context of the cosmological constant seesaw. In section 3 we discuss specific models of Universe decay such as chain inflation and string theory which can be used to calculate the decay constant $\Gamma$.  In section 4 we discuss how third quantization and Universe entanglement can incorporated into  the cosmological constant seesaw in a simplified context of chain inflation involving a four form flux. In section 5 we discuss the relation of the mixing parameter to various approaches to the hierarchy problem. In section 6 we describe the main conclusions of the paper. 

\section{Universe decay and the cosmological constant seesaw}

Consider a simple ansatz where the metric is of the form:
\[
ds^2 = -N^2dt^2 + a^2( e^{2\beta_1}dx^2 + e^{2\beta_2} dy^2 + e^{2\beta_3}dz^2)
\]
and $N$, $a$ and $\beta_i$ are all functions of $t$. The variables $\beta_i$ satisfy the relation:
\[
\beta_1 + \beta_2 + \beta_3 = 0
\]
This constraint is satisfied by defining new variables $\beta_-$ and $\beta_+$ \cite{Misner:1969hg}\cite{Kuchar:1989tj}\cite{Erickson:2003zm} through:
\[
\begin{array}{l}
 \beta _1  = \beta _ +   + \sqrt 3 \beta _ -   \\
 \beta _1  = \beta _ +   - \sqrt 3 \beta _ -   \\
 \beta _3  =  - 2\beta _ +   \\
 \end{array}
\]
Experimentally there is some evidence that the $\beta_i$ variables are nonzero \cite{Campanelli:2006vb}. If true this  means that the Universe has has a slightly oblong shape. Analysis of quadrapole density perturbations indicate \cite{Campanelli:2006vb}. 
\[
\begin{array}{l}
 \beta _1  = (5 \times 10^{ - 5} )\frac{1}{3} \\
 \beta _2  = (5 \times 10^{ - 5} )\frac{1}{3} \\
 \beta _3  = ( - 5 \times 10^{ - 5} )\frac{2}{3} \\
 \end{array}
\]
For our purpose we simply use the $\beta_i$ variables as a convenient parameter space in which to study Universe decay.

The Friedman equation with vacuum energy density and radiation density  is given by:
\[
3m_{P}^2 \frac{1}{{N^2 }}\left( {\frac{{\dot a}}{a}} \right)^2  = \lambda  - m_{P}^2 \frac{k}{{a^2 }} + \frac{{N_{rad} }}{{a^4 }} + \frac{{P_ + ^2  + P_ - ^2 }}{{a^6 }}
\]
One can also add a matter term to the Friedmann equation of the form $\frac{N_{matt}}{a^3}$ however as one goes to early times (smaller $a$) the radiation term is more important than the matter term. In the above $P_{\pm}$ are canonically conjugate to the $\beta_{\pm}$ variables.
At very early times with large $\lambda$ or very late times with small $\lambda$ the cosmological constant term dominates all other terms. In the $N=1$ gauge the solution is then:
\begin{equation}
a(t) = a(0)e^{tm_{P}^{ - 1} \sqrt {\lambda /3} }
\end{equation}
When curvature is present the classical dynamics of the $a, \beta_i$ variables is chaotic and described by the motion of a point in $a, \beta_+, \beta_-$  space bouncing off effectively triangular walls induced from a effective potential for the $\beta_i$. This effective potential comes from the spatial curvature. In this paper we shall mostly take  $k=0$ and no spatial curvature. This is consistent with recent data $k = H^2a^2(\Omega_0-1)$ where $H$ is the Hubble parameter and $\Omega_0 = 1.02 \pm .02$ \cite{Reboucas:2006ri}\cite{Adler:2005mn}.

One thing that doesn't happen classically is that $\lambda$ does not change and Universe decay does not take place. In particular although the solution to the Friedman equation involves DeSitter expansion, this expansion does not end. So there is no exit from the inflationary phase in this model classically. This point is addressed in section 2 of reference \cite{Watson:2006px}.

Quantum mechanically it is  less clear that one can cannot exit from the inflationary phase. For example quantum tunneling might occur. However to exit from the inflationary phase one needs a way to change the cosmological constant after some period of large DeSitter expansion to an extremely fine tuned small value that we see today. We shall show that the cosmological constant seesaw together with Universe decay can accomplish this.

First writing the Friedmann equation in terms of canonical momentum we have:
\[
\frac{{\bar{m}_{P}^{ - 2} }}{{9a^4 }}P_a^2  = \lambda  - m_{P}^2 \frac{k}{{a^2 }} + \frac{{N_{rad} }}{{a^4
}} + \frac{{P_ + ^2  + P_ - ^2 }}{{a^6 }}
\]
Here we have defined:
\[
\bar m_P^2 = 27(2\pi)^3m_P^2
\]
The Friedmann equation is a  classical equation. We don't expect it to be valid quantum mechanically. One way to turn the Friedmann equation  into a statement consistent with quantum mechanics is to introduce the concept of a wave function of the Universe $\psi$ and write the WDW equation:
\[
\frac{{\bar m_{P}^{ - 2} }}{{9a^4 }}P_a^2 \psi  = (\lambda  - m_{P}^2 \frac{k}{{a^2 }} + \frac{{N_{rad} }
}{{a^4 }} + \frac{{P_ + ^2  + P_ - ^2 }}{{a^6 }})\psi
\]
This is second order differential equation. Taking the square root of the differential operator we have the first order equation:
\[
\bar  m_{P}^{ - 1} i\partial _{a^3 } \psi  = (\lambda  - m_{P}^2 \frac{k}{{a^2 }} + \frac{{N_{rad}
}}{{a^4 }} + \frac{{P_ + ^2  + P_ - ^2 }}{{a^6 }})^{1/2} \psi
\]
For large $\lambda$ or for small $\lambda$ and large radius the cosmological constant dominates and we have the solution \cite{Vilenkin:1994rn}:
\begin{equation}
\psi  = C(a)\exp ( - i\sqrt \lambda  a^3 \bar{m}_{P} )
\end{equation}
where $C(a)$ is a coefficient that depends on normal ordering. This wave function corresponds to the classical solution (2.1).

The operation of taking the square root is better done using a Dirac square root method in which case the WDW equation is written:
\[
 \bar  m_{Pl}^{ - 1} \left( {\begin{array}{*{20}c}
   {i\partial _{a^3 } } & {a^{ - 3} P}  \\
   {a^{ - 3} P^ *  } & {i\partial _{a^3 } }  \\
\end{array}} \right)\psi  = (\lambda  - m_{P}^2 \frac{k}{{a^2 }} + \frac{{N_{rad} }}{{a^4 }})^{1/2} \psi
\]
where we have defined $P = P_+ + iP_-$. 

In this form it is easiest to introduce the Matrix square root of $\Lambda$ and the method of treating $\Lambda$ as a $2\times 2$ matrix. Working for large $\lambda$ or for small $\lambda$ and large radius we can omit the $N_{rad}$ and $P$ terms in the equation to obtain:
\[
\begin{array}{l}
 \bar m_{P}^{ - 1} i\partial _{a^3 } \psi _1  = 0\psi _1  + c_{12}m_{EW}^2 \psi _2   \\
  
 \bar m_{P}^{ - 1} i\partial _{a^3 } \psi _2  = c_{22}m_P^2 \psi _2  + c_{12} m_{EW}^2\psi _1  \\
   \end{array}
\]
 The WDW equations for the eigenmodes $\psi_{I,II}$ of the cosmological constant matrix are very simple and are given by:
\[
\begin{array}{l}
 \bar m_P^{-1} i\partial _{a^3 } \psi _I  =  - \sqrt {\lambda _I } \psi _I  \\
 \bar m_P^{-1} i\partial _{a^3 } \psi _{II}  = \sqrt {\lambda _{II} } \psi _{II}  - i\frac{\Gamma }{2\bar m_P^2}\psi _{II}  \\
 \end{array}
\]
We have introduced the $\Gamma$ parameter which describes the instability of the Universe with the large eigenvalue $\lambda_{II}$. We discuss this parameter in more detail in the next section. The solutions to these eigenmode equations are
\[
\begin{array}{l}
 \psi _I  = e^{i\sqrt {\lambda _I } a^3 \bar m_P }  \\
 \psi _{II}  = e^{ - i\sqrt {\lambda _{II} } a^3 \bar m_P } e^{ - \frac{\Gamma }{2}a^3/ \bar m_P }  \\
 \end{array}
\]
The solutions to the coupled WDW equations are:
\[
\begin{array}{l}
 \psi _1  = \frac{1}{{\sqrt {\lambda _{I}^{1/2}  + \lambda _{II}^{1/2} } }}(\lambda _{II}^{1/4} \psi _I  + \lambda _I^{1/4} \psi _{II} ) \\
 \psi _2  = \frac{1}{{\sqrt {\lambda _{I}^{1/2}  + \lambda _{II}^{1/2} } }}(- \lambda _I^{1/4} \psi _I  + \lambda _{II}^{1/4} \psi _{II} ) \\
 \end{array}
\]

From the fact that the eigenvalue of the operator $P_a$ on $\psi_1$ is negative at large $a$ we see that the wave function $\psi_1$ describes a contracting Universe at large radius. We can use the fact that complex conjugation on a quantum mechanical wave function $\psi_c = \psi^*$ will map a contracting to an expanding Universe \cite{Vilenkin:1994rn}\cite{Buonanno:1996um}
 to identify wave functions describing accelerating and inflating phases of the Universe as follows:
\[
\begin{array}{l}
 \psi _{accel}  = \frac{1}{{\sqrt {\lambda _{I}^{1/2} + \lambda _{II}^{1/2} } }}(\lambda _{II}^{1/4} \psi _I^ *   + \lambda _I^{1/4} \psi _{II}^ *  ) \\
 \psi _{infl}  = \frac{1}{{\sqrt {\lambda _{I}^{1/2}  + \lambda _{II}^{1/2} } }}(  -\lambda _I^{1/4} \psi _I  + \lambda _{II}^{1/4} \psi _{II} ) \\
 \end{array}
\]
Substituting the expression for the eigenvalue and eigenvectors this reduces to:
\[
\begin{array}{l}
 \psi _{accel}  = e^{-i\frac{{c_{12}^2 m_{EW}^4 }}{{c_{22} m_P^2 }}a^3 \bar m_P }  + \frac{{c_{12} m_{EW}^2 }}{{c_{22} m_P^2 }}e^{ ic_{22} m_P^2 a^3 \bar m_P } e^{ - \frac{\Gamma }{2}a^3/\bar m_P  }  \\
 \psi _{infl}  =  - \frac{{c_{12} m_{EW}^2 }}{{c_{22} m_P^2 }}e^{i\frac{{c_{12}^2 m_{EW}^4 }}{{c_{22} m_P^2 }}a^3 \bar m_P }  + e^{ - ic_{22} m_P^2 a^3  \bar m_P } e^{ - \frac{\Gamma }{2}a^3/\bar m_P }  \\
 \end{array}
\]
which is the result alluded to in the introduction. The important point is that applying quantum mechanics to gravity allows one to introduce a dynamical set of equations that behave as an strongly inflating Universe at early times and an accelerating Universe at late times. The large DeSitter expansion part of the wave function (the second term in $\psi_{accel}$) dies out after a certain period determined by the inverse of the parameter $\Gamma$. We interpret this  as the end of inflation. The novel feature is the quantum entanglement of states with two different cosmological constants. It is this feature which differs most strongly the classical cosmology which always has a single cosmological constant.

We define the mixing angle $\theta$ by:
\[
\begin{array}{l}
 \sin \theta  = \frac{{\lambda _I^{1/4} }}{{\sqrt {\lambda _I^{1/2}  + \lambda _{II}^{1/2} } }} \approx \frac{{c_{12} m_{EW}^2 }}{{c_{22} m_P^2 }} \\
 \cos \theta  = \frac{{\lambda _{II}^{1/4} }}{{\sqrt {\lambda _I^{1/2}  + \lambda _{II}^{1/2} } }} \approx 1 - \frac{1}{2}\frac{{c_{12}^2 m_{EW}^4 }}{{c_{22}^2 m_P^4 }} \\
 \end{array}
\]
The mixed wave functions are then given by:
\[
\begin{array}{l}
 \psi _{1}  = \cos \theta \psi _I  + \sin \theta \psi _{II}  \\
 \psi _{2}  =  - \sin \theta \psi _I  + \cos \theta \psi _{II}  \\
 \end{array}
\]
Because the mixing angle $\theta$ is so small $\psi_1$ is nearly a purely accelerating Universe with cosmological constant $\lambda_I$ while $\psi_2$ is nearly a purely inflating Universe with cosmological constant $\lambda_{II}$.
The transition amplitude to start with the inflationary state $\psi_2$ and end up with the accelerating state $\psi_1$ is then:
\[
\left\langle {\psi _1 } \right|{\psi _2 ,a^3 } \rangle  = \sin \theta \cos \theta (e^{ - i\sqrt {\lambda _{II} } a^3 \bar m_P } e^{ - \Gamma a^3 /2\bar m_P}  - e^{ - i\sqrt {\lambda _I } a^3 \bar
m_P } )
\]
the notation indicates that we are using the $\psi_2$ wave function when the Universe had a volume $a^3$. The transition probability is then:
\[
P(2 \to 1;a^3 ) = \sin ^2 \theta \cos ^2 \theta (1 - 2e^{ - \Gamma a^3 /2\bar m_P} \cos(( \sqrt {\lambda _{II} }  - \sqrt {\lambda _I }  )a^3 \bar m_P  ) + e^{ - \Gamma a^3/\bar m_P} )
\]
Similarly one can also define transition amplitudes to start with the inflationary state and transit into cosmological constant eigenstates $\psi_I$ or $\psi_{II}$. These are
\begin{equation}
\begin{array}{l}
 \left\langle {\psi _I } \right|{\psi _2 ,a^3 } \rangle  =  - \sin \theta e^{ - i\sqrt {\lambda _I } a^3 \bar m_P }  \\
 \left\langle {\psi _{II} } \right|{\psi _2 ,a^3 } \rangle  = \cos \theta e^{ - i\sqrt {\lambda _{II} } a^3 \bar m_P } e^{ - \Gamma a^3 /2\bar m_P}  \\
 \end{array}
\end{equation}
and the transition probabilities are given by
\[
\begin{array}{l}
 P(2 \to I;a^3 ) = (\sin \theta )^2  \\
 P(2 \to II;a^3 ) = (\cos \theta )^2 e^{ - \Gamma a^3/\bar m_P }  \\
 \end{array}
\]
While both of these are very small, the relative probability for transition into the eigenfunction with the small cosmological constant eigenvalue dominates once the Universe expands. This is  because of the probability to be in the state with  the large eigenvalue is suppressed exponentially by the factor $e^{ - \Gamma a^3/\bar m_P }$ with decay parameter $\Gamma$.

The solution we have found to the WDW equation with mixing and decay is analogous to the analysis of Gell-Mann and Pais of the $K_L$ and $K_S$ Kaon mixing and decay system \cite{Gell-Mann:1955jx}\cite{Feynman:1965kb}. Like that analysis we can describe certain general features of the quantum system in terms of three parameters $c_{12}$, $c_{22}$ and $\Gamma$. In the Kaon analogy the short lived Kaon $K_S$ is analogous to the decaying inflationary wave function. The long lived Kaon $K_L$ is analogous to the stable or metastable accelerating wave function which describes our Universe now. To go further one needs specific  models of Universe decay some of which we describe in the next section. This is analogous to the need for a more fundamental description of Kaon interactions to derive the parameters the Kaon mass matrix.

In the above we used to solution for  zero $\beta_i$ for simplicity. The solution to the WDW equations with nonzero $\beta_i$  can also be written 
down. The main difference is a more complicated form of the eigenfunctions. Instead of a simple exponential in (2.2) one finds a Bessel function solution \cite{McGuigan:1989yb}:
\[
\psi = J_{ - i\left| P \right|} (\sqrt \lambda  a^3 \bar m_P )e^{iP_ +  \beta _ +   + iP_ -  \beta _ -  }
\]
which is an eigenmode of the WDW equation:
\[
 - \frac{1}{{\bar m_P^2 a^6 }}\partial _{a^3 } (a^6 \partial _{a^3 } \psi ) = (\frac{{\left| P \right|^2 }}{{\bar m_P^2 a^6 }} + \lambda )\psi
\]
The above solutions involve a specific choice of operator ordering. Other choices lead to different values of $C(a)$ in the the large $a$ expansion of the wave function (2.2).

For the general case with nonzero $\beta_i$, $N_{rad}$ and $N_{matt}$ one can use the WKB form of the solutions given by:
\[
\psi = \exp \left( { - i\bar m_P \int_{a_0^3 }^{a^3 } {d\bar a^3 (\lambda  + \frac{{\left| P \right|^2 }}{{\bar m_P^2 a^6 }} + \frac{{N_{rad} }}{{a^4 }}}  + \frac{{N_{matt} }}{{a^3 }} - \frac{{\bar m_P^2 k}}{{a^2 }})^{1/2} )} \right)e^{iP_ +  \beta _ +   + iP_ -  \beta _ -  }
\]
Finally for the special case where $k=1$ and $\beta_i=0$, $N_{rad}$, $N_{matt}$ all zero one can express the solutions to the WDW equation in terms of Airy functions
\cite{Duncan:1990fr}\cite{Vilenkin:1994rn}.

\section{Universe decay and quantum tunneling}

One natural approach to interpret the parameter $\Gamma$ in the cosmological constant seesaw is quantum tunneling. The parameter $\Gamma$ in that context is usually expressed as:
\[
\Gamma  = Ae^{ - B}
\]
where $B$ is a decay exponent and $A$ is a prefactor. The form of $B$ depends on the specific model of tunneling in Universe decay. 

Tunneling rates are very important in inflationary models. In the old inflationary models single tunneling events achieved sufficient inflation but where incompatible with thermalization and a graceful exit from inflation. In new inflationary models slow roll potentials or a sequence of tunneling events, chain inflation \cite{Freese:2006fk},  are used have sufficient inflation to solve cosmological problems while still satisfying graceful exit conditions. 
We discuss some treatments of quantum tunneling using field theory, chain inflation and string theory below. The expressions of the tunneling rate $\Gamma$ are then included into the WDW semiclassical analysis in the following section.

\subsection{Tunneling in field theory}
 
For scalar field theories at zero temperature the decay exponent can be written as \cite{Sher:1988mj}:
\begin{equation}
B = 2\pi ^2 \int_0^\infty  {\rho ^3 d\rho (\frac{1}{2}} \left( {\frac{{d\phi }}{{d\rho }}} \right)^2  + U)
\end{equation}
Where $U$ is an $O(4)$ invariant scalar potential obeying the second order differential equation:
\[
\frac{{d^2 \phi }}{{d\rho^2 }} + \frac{3}{\rho}\frac{{d\phi }}{{d\rho}} = \frac{{dU}}{{d\phi }}
\]
For scalar field theories at nonzero temperature $T$ the decay exponent is \cite{Sher:1988mj}:
\[
B = \frac{{4\pi }}{T}\int_0^\infty  {r^2 dr(\frac{1}{2}} \left( {\frac{{d\phi }}{{dr}}} \right)^2  + V)
\]
where $V$ is the $O(3)$ invariant scalar potential obeying the second order differential equation:
\[
\frac{{d^2 \phi }}{{dr^2 }} + \frac{2}{r}\frac{{d\phi }}{{dr}} = \frac{{dV}}{{d\phi }}
\]
For a quantum field generalization of the square barrier potential the decay exponent can be computed exactly and gives \cite{Duncan:1992ai}:
\[
B = 2\pi ^2 \frac{{\left( {\phi _I  - \phi _{II} } \right)^4 }}{{\left( {\left( {V_{barrier}  - V_I } \right)^{1/3}  - \left( {V_{barrier}  - V_{II} } \right)^{1/3} } \right)^3 }}
\]
For a quantum field generalization of a triangular barrier potential the decay exponent can also be evaluated exactly to give \cite{Duncan:1992ai}: 
\[
B = \frac{{2\pi ^2 }}{3}\frac{{\left( {(\phi _{barrier}  - \phi _I )^2  - (\phi _{barrier}  - \phi _{II} )^2 } \right)^2 }}{{V_{barrier}  - V_{II} }}
\]
In these formulae $V_{barrier}$ and $\phi_{barrier}$ are the height  and field value at the  barrier potential. $V_{I,II}$ and $\phi_{I,II}$ are the height and field values to the left and right of the barrier. For more realistic potentials the decay exponent can be evaluated numerically using (3.1).

Another example of Universe decay in field theory is the transition from a Kaluza Klein (KK) vacuum of spatial topology $R^3\times S^1$ to an uncompactified vacuum with spatial topology $R^4$. The decay exponent was calculated in \cite{Witten:1981gj} with the value:
\[
B = \pi \frac{{R^2 }}{{4G}} = 2\pi ^2 m_P^2 R^2
\]
where $R$ is the compactified radius. The KK vacuum can be stabilized if suitable fermionic matter is present \cite{Rubin:1983zz}. Familiar physical transitions can sometimes be cast in the form of vacuum decay. One example of this is the formation of a small black hole from the quantum instability of hot flat space. The decay exponent in this case is estimated to be \cite{Gross:1982cv}:
\[
B = \frac{1}{{16\pi GT^2 }} = \frac{1}{2}\frac{{m_P^2 }}{{T^2 }}
\]
where $T$ is the temperature of hot flat space. Interestingly this type of instability can be investigated using the mathematical techniques of Ricci flow with surgery \cite{Headrick:2006ti}. The relation of this mathematical technique to a renormalization group sigma model approach to string theory is a subject of great current interest \cite{Bakas:2006bz}\cite{Tseytlin:2006ak}.

\subsection{Chain inflation}

Chain inflation \cite{Freese:2006fk} is a new approach to Universe decay involving a four form flux in which the decay proceeds through a series of discrete steps labeled by a index $n$ each with it's own decay constant $B_n$. In this approach one models the dependence of the cosmological constant on a four form flux $Q$ as:
\[
\lambda (Q) = c_0  + c_1 Q^2  + c_2 Q^4  +  \ldots
\]
The flux is quantized in string models and can be parametrized as:
\[
Q = qn
\]
where $n$ is an integer. The change in the vacuum energy from one step in the chain to an adjacent step is given by:
\[
\lambda (q(n - 1)) - \lambda (qn) =  - c_1 q^2 (2n - 1) - c_2 q^4 (4n^3  - 6n^2  + 4n - 1) +  \ldots
\]
The decay exponent in chain inflation can be estimated in a thin wall approximation as:
\[
B_n= \frac{{27\pi ^2 }}{2}\frac{{\tau ^4 }}{{\left| {\lambda (q(n - 1)) - \lambda (qn)} \right|^3 }} = \frac{{27\pi ^2 }}{8}\frac{{m_P^4 q^4 }}{{\left| {\lambda (q(n - 1)) - \lambda (qn)} \right|^3 }}
\]
where $\tau  = \frac{1}{{\sqrt 2 }}qm_P$. The decay parameter $\Gamma(Q)$ is then of the form:
\begin{equation}
\Gamma(Q) = Ae^{-B_n}
\end{equation}
in later sections we find it convenient to divide this by $\bar m_P$ to obtain a quantity of dimension $mass^3$. 

The final state after chain inflation is not known. The decay rate for the transitions may slow to a metastable state or one could end  up in a final state with vacuum energy $c_0$.
In \cite{Freese:2006fk} a chain evolution was described  from an initial to final state with vacuum energies:
\[
\begin{array}{l}
 \lambda _{initial}  = 10^{ - 4} m_P^4  \\
 \lambda _{final}  = 10^{ - 80} m_P^4  = (24.53MeV)^4  \\
 \end{array}
\]
Although this represents a drop in many orders of magnitude from the initial value this is still not small enough to approach the observed value of the vacuum energy. Another important point is that the dependence of the cosmological constant on the flux can be quite complex nonperturbatively as illustrated in the Matrix model calculations of 2d string theory \cite{Maldacena:2005he}. 

If the changes in vacuum energy between transitions are small enough, the decrease can be modeled using a time dependent vacuum energy as in the inflatonless inflation approach of \cite{Watson:2006px}. There they find that the vacuum energy depends on time as:
\[
\begin{array}{l}
 \lambda (t) = \lambda _0 e^{ - \Gamma t}  \\
 \Delta \lambda (t) =  - \Gamma \Delta t\lambda (t) \\
 \end{array}
\]
In the above one divides $\Gamma$ by $\bar m_P^3$ to obtain a quantity of the dimension $mass$. Recent data indicates that whatever is giving rise to dark energy hasn't changed much in the history of the Universe \cite{Riess:2006fw}. This may constrain time dependent vacuum energy approaches. We discuss further implications of chain inflation and the cosmological constant seesaw in section 4.

\subsection{Tunneling in string theory}

One often hears the statement that the observed values of the cosmological constant is 122 orders of magnitude smaller than the expected value of $m_{P}^4$. In a field theoretic context this is not really correct. This is because in field theory one usually assumes a cutoff regularization and calculates $\lambda$ of order $M_{cutoff}^4$. If one removes this cutoff this simply diverges so  the expected value is actually infinity not something of the order of $m_P^4$. This creates a problem for the implementation of the cosmological constant seesaw as one actually wants to make use of a large value of the cosmological constant of the order of $m_P^4$ to define the quantity $\lambda_{II}$. One then wants to use this large eigenvalue to describe inflation and the small eigenvalue to describe the late time accelerating Universe with eigenvalue of order $\frac{m_{EW}^8}{m_P^4}$.

In string models one can give a rather explicit representation of the cosmological constant as a sum over two dimensional world sheets. On can even use large $N$ methods to define the cosmological constant in string theory nonperturbatively. This allows one to give finite values to the parameters of the cosmological constant seesaw whereas for typical  field theories these would be divergent quantities (One exception may be $N=8$ supergravity but in that case the vacuum energy vanishes). 

Tunneling using moduli potentials in string theory offers an improvement over field  theory calculations because  the initial and final vacuum energies are finite and well defined. For example consider the $O(16)\times O(16)$ string compactified on $T^2$. Vacuum energies of two local minima and one local maximum of the moduli potential are given by \cite{Ginsparg:1986wr}\cite{Nair:1986zn}
:
\[
\begin{array}{l}
V_I = \lambda _{E_8\times E_8}  = 0 \\
 V_{II} = \lambda _{O(16)\times O(16) \times SU_3}  = .93 c_{22}^2 \frac{.0371}{(2\pi \alpha ' )^4 }\\
 V_{barrier} = \lambda _{O(16)\times O(16) \times SU_2 \times SU_2}  = c_{22}^2 \frac{.0371}{(2\pi \alpha ' )^4}\\
 \end{array}
\]
In the above list we also include the $E_8\times E_8$ string which has zero vacuum energy and is supersymmetric. Blum and Dienes have found a continuous connection between the $E_8\times E_8$  theory and the $O(16)\times O(16)$ compactified string \cite{Blum:1997gw}\cite{Blum:1997cs}. The subscripts $SU(2)\times SU(2)$ and $SU(3)$ indicate an enhanced symetry group that occurs at the corresponding points in moduli space. The parameter $c_{22}^2$ is used to normalize the moduli potential $V_{II}$ with respect to the ten dimensional vacuum energy of the $O(16)\times O(16)$  uncompactified string which is $.0371/(2\pi \alpha ' )^5$\cite{Ginsparg:1986wr}. Dilaton tadpole infared divergences for the compactified $O(16)\times O(16)$ theory are troublesome but can be removed by using the Fischler-Susskind mechanism  \cite{Fischler:1986ci}\cite{Fischler:1986tb} and including the effect of the one-loop cosmological constant on the tree level equations of motion \cite{Callan:1986bc}. The $O(16) \times O(16)$ model represents a simplified context to study tunneling associated with moduli fields. More realistic discussions of moduli potentials, stabilization  and inflation in string theory is found in \cite{Berg:2004ek}.

Recently there has been some interest in nonsupersymmetric string compactifications \cite{Dienes:2006ut}\cite{Dine:2006gx}. To this end one  can replace the $T^2$ with $K=\Sigma_g\times S^4$ where $\Sigma_g$ is a genus $g$ surface. These compactifications of the $O(16)\times O(16)$ string yield an $SO(10)$ grand unified group with chiral matter and a number of generations $N_{gen} = \left| \chi(K) \right|/2$ . Using direct sums one can form three generation models with $K$ given by \cite{Jackiw:1986pg}\cite{Green:1987mn}:
\[
\begin{array}{l}
 (S^2  \times S^4 ) \oplus (S^2  \times S^4 ) \\
 (T^2  \times S^4 ) \oplus (\Sigma _2  \times S^4 ) \\
 (S^2  \times S^4 ) \oplus (\Sigma _3  \times S^4 ) \\
 (T^2  \times S^4 ) \oplus (T^2  \times S^4 ) \oplus (T^2  \times S^4 ) \oplus (T^2  \times S^4 ) \\
 \end{array}
\]
Compactified spaces such as $\Sigma_2\times S^4$ that are nonsimply connected are particularly interesting as the $SO(10)$ group can be further broken through the use of Wilson lines. Although these models are not viable as GUT models because of the large cosmological constant, they may find a place in the cosmological constant seesaw scenario as states associated with the large cosmological constant eigenvalue. 

\section{Third quantization and Universe entanglement}

Recall that there are three fairly distinct application areas of string models (besides mathematics):

1) Grand unification and gravity.

2) String gauge theory correspondence.

3) 2d quantum gravity model of topology change and quantum cosmology.

While some have questioned the quantitative relevance of 1) and 2) \cite{Woit:2006js}\cite{McLerran} few people who have thought about nonperturbative quantum gravity would say that 3) did not yield some thought provoking insights \cite{Polchinski:1989fn}\cite{Cooper:1991vg}\cite{DaCunha:2003fm}. Recall that the first large N Matrix string correspondence occurred in studies of nonperturbative 2d quantum gravity. The ideas of many Universe Hilbert space and third quantization also came from the string field theory approach to the 2d world sheet. 

Of course as with any toy model 3) has  limitations (no gravitons in 2d for example) but the third quantized 2d gravity approach furnishes at least one answer to the  question "What is string theory?" \cite{Polchinski:1994mb} albeit only for specific backgrounds. In physics extended objects are typically realized as being made of something (protons are made up of quarks and gluons for example) so a natural question is what are strings made of? The 2d gravity approach answers this question by saying they are spacial slices of a 2d spacetime in 2d quantum gravity. One doesn't need to ask what empty space is made of (e.g. mechanical models of the aether) so the conceptual question of what strings are made of is handled in a simple geometrical way in the 2d gravity approach. 
  
The third quantized baby Universe ideas which originally came from attempts to generalize 3) to higher dimensions have recently resurfaced in new approaches to exact wave functions of the Universe calculated from string theory. When gravity backgrounds have duals described by free fermion Fock spaces the wave function of the Universe can be shown to be a coherent sum of contributions which corresponds to a  ensemble of gravity duals \cite{Brill:1991rw}\cite{Rey:1998yx}\cite{Maldacena:2004rf}\cite{Dijkgraaf:2005bp}\cite{Aganagic:2006je}. Specific examples in \cite{Aganagic:2006je} are given in which  one has a fermionic description of a large N 2d YM holographic dual to an ensemble of manifolds of the form $AdS_2\times S^2 \times CY_6$ with different fluxes. Because of the coherent sum these are said to be examples of baby Universe entanglement.

One of the simplest contexts in which to introduce Universe entanglement is 4d gravity coupled to a four form. In this case the Friedmann equation is:
\[
3m_P^2\frac{{\dot a^2 }}{{N^2 a^2 }} = \lambda  + \frac{{\dot A^2 }}{{N^2 a^6 }}
\]
Here we define the  quantity $A = \int A_{ijk}dx^i dx^j dx^k$ which is canonically conjugate to the four form charge. The WDW equation is:
\[
 - \bar m_P^{ - 2} \frac{{\partial ^2 }}{{\partial (a^3 )^2 }}\psi  = (\lambda  + Q^2 )\psi
\]
Where $Q$ is the four form flux. The simplicity comes from the fact that a four form field strength has no physical degrees of freedom in 4d. In fact the WDW equation is the same as for ordinary 4d gravity with a $Q$ dependent cosmological constant. In string theory this $Q$ dependence can be quite intricate. Here we consider only the quadratic dependence coming from the tree level Lagrangian.

The WDW equation associated with chain inflation with four form flux $Q$ has the WKB solution:
\[
\psi_Q = e^{ - i\sqrt {Q^2  + \lambda } a^3 \bar m_P } e^{iQA}
\]
One can also form superposition of such solutions given by:
\[
\psi (a,A) = \sum\limits_Q {(\alpha_Q } e^{ - i\sqrt {Q^2  + \lambda } a^3 \bar m_P }  + \alpha_{ - Q}^+  e^{i\sqrt {Q^2  + \lambda } a^3 \bar m_P } )e^{iQA}
\]
 As a wave function of this form is a linear combination wave functions associated with $\lambda_Q$ we say that this is an entangled state of Universes each with a different value of $\lambda_Q=\lambda+Q^2$. If one thinks of $\lambda_Q$ as a cosmological constant then a state of this form has no definite cosmological constant, although it is still physical and satisfies the WDW equation. It is probably better to think of $Q$ as a canonical momentum in the space of three forms $A_{123}$.  In that case one can define a cosmological constant associated with this superimposed state as $\lambda(Q=0) = \lambda$. This is analogous to how one can define mass from $E^2 = P^2 + M^2$ evaluated at $P=0$. It is easy to incorporate the quantization of fluxes into the wave function. This follows if the theory is periodic in the $A$ variable.

Our previous treatment of the cosmological constant seesaw was based on an application of the quantum mechanical treatment of mixing similar to that used by Gell-Mann and Pais for the Kaon system. A quantum field description of mixing has been developed in \cite{Blasone:2001du}\cite{Capolupo:2004av}. This gives a modification to the mixing formula associated with the fact that the wave function is treated as an operator and is second quantized. To apply this  quantum field formalism to the cosmological constant seesaw and the WDW equation one third quantizes the theory.  

Third quantization is introduced into the superposition wave function by letting the $\alpha_Q$ coefficients become operators which satisfy the commutation relations:
\[
[\alpha _Q ,\alpha _{Q'}^ +  ] = \delta _{Q,Q'}
\]
The eigenfunctions then become operators of the form:
\[
\begin{array}{l}
 \psi _I (a,A) = \sum\limits_Q {(\alpha _{Q,I} } e^{ - i\sqrt {Q^2  + \lambda _I } a^3 \bar m_P }  + \alpha _{ - Q,I}^ +  e^{i\sqrt {Q^2  + \lambda _I } a^3 \bar m_P } )e^{iQA} e^{ - \Gamma _I (Q)a^3 /2}  \\
 \psi _{II} (a,A) = \sum\limits_Q {(\alpha _{Q,II} } e^{ - i\sqrt {Q^2  + \lambda _{II} } a^3 \bar m_P }  + \alpha _{ - Q,II}^ +  e^{i\sqrt {Q^2  + \lambda _{II} } a^3 \bar m_P } )e^{iQA} e^{ - \Gamma _{II} (Q)a^3 /2}  \\
 \end{array}
\]
In the above we also include the decay constants of chain inflation ($\Gamma(Q)$ from formula (3.2)) in the wave function of the Universe. The mixed operators are defined in terms of the above expansion and are given by:
\[
\begin{array}{l}
 \alpha _{Q,1} (a^3 ) = \cos \theta \alpha _{QI} e^{ - \Gamma _I (Q)a^3 /2}  + \sin \theta (u_Q^* (a^3 )\alpha _{Q,II}  + v_Q (a^3 )\alpha _{ - Q,II} )e^{ - \Gamma _{II} (Q)a^3 /2}  \\
 \alpha _{Q,2} (a^3 ) = \cos \theta \alpha _{Q,II} e^{ - \Gamma _{II} (Q)a^3 /2}  - \sin \theta (u_Q (a^3 )\alpha _{Q,I}  + v_Q^* (a^3 )\alpha _{ - Q,I}^ +  )e^{ - \Gamma _I (Q)a^3 /2}  \\
 \end{array}
\]
where we have used the definition of the mixing angle $\theta$ from section 2 and have defined:
\[
\begin{array}{l}
 u_Q (a^3 ) = \frac{1}{2}\left( {\sqrt {\frac{{\Omega _{QI} }}{{\Omega _{QII} }}}  + \sqrt {\frac{{\Omega _{QII} }}{{\Omega _{QI} }}} } \right)e^{i(\Omega _{QII}  - \Omega _{QI} )a^3 }  \\
 v_Q (a^3 ) = \frac{1}{2}\left( {\sqrt {\frac{{\Omega _{QI} }}{{\Omega _{QII} }}}  - \sqrt {\frac{{\Omega _{QII} }}{{\Omega _{QI} }}} } \right)e^{i(\Omega _{QII}  + \Omega _{QI} )a^3 }  \\
 \end{array}
\]
as well as
\[
\begin{array}{l}
\Omega _{QI}  = \bar m_P\sqrt {Q^2  + \lambda _{I} } \\
\Omega _{QII}  = \bar m_P\sqrt {Q^2  + \lambda _{II} } \\
\end{array}
\]
The third quantized mixing probability is analogous to the quantum field mixing equations in \cite{Blasone:2001du}\cite{Capolupo:2004av}. They are given by:
\[
 P(2 \to 1;a^3 ,Q) = \left| {[\alpha _{Q,2} (a^3 ),\alpha _{Q,1}^ +  (0)]} \right|^2  - \left| {[\alpha _{ - Q,2}^ +  (a^3 ),\alpha _{Q,1}^ +  (0)]} \right|^2  
 \]
Using the commutation relations this becomes:
\begin{equation}
\begin{array}{l}
 P(2 \to 1;a^3 ,Q) = \\
(\sin \theta \cos \theta )^2 (e^{ - \Gamma _I (Q)a^3 }  + e^{ - \Gamma _{II} (Q)a^3 }  - 2e^{ - (\Gamma _I (Q) + \Gamma _{II} (Q))a^3 /2} \cos ((\Omega _{QI}  - \Omega _{QII}
)a^3 )) \\
  - 4\left| {v_Q } \right|^2 (\sin \theta cos \theta )^2 \sin (\Omega _{QI} a^3 )\sin (\Omega _{QII} a^3 ) e^{ - (\Gamma _I (Q) + \Gamma _{II} (Q))a^3 /2}\\
  \\
 \end{array}
\end{equation}
The third quantized modification to the usual mixing formula is proportional to the quantity $v_Q^2$ in the last term. This takes a maximum value given by:
\[
v_{\max }^2  = \frac{1}{4}\frac{{(\lambda _{II}^{1/2}  - \lambda _I^{1/2} )^2 }}{{\lambda _I^{1/2} \lambda _{II}^{1/2} }} \approx \frac{{c_{22}^2 m_P^4 }}{{4c_{12}^2 m_{EW}^4 }}
\]
This is extremely large and effectively cancels the small factor $(\sin \theta )^2$ in the last term of (4.1). Thus the third quantized description of Universe mixing in the cosmological constant seesaw is characterized by a very small mixing angle and a very large third quantized modification to the quantum mixing equation. This is consistent with previous studies which found large third quantized corrections to uncertainty relations in the early Universe  \cite{Abe:1993ye}\cite{Horiguchi:1993tu}\cite{Pohle:1991jz}. The quantum mixing formula provides a bridge between early Universe phenomena and the low energy experimental measurement of the cosmological constant. In the Kaon analogy the Kaon mixing provides a bridge between low energy meson wave functions and high mass states like the $W$ boson or top quark. The cosmological constant seesaw may provide a similar high energy/low energy relation for the gravitational interactions.

\section{Mixing angle and the hierarchy problem}

Although there are many approaches to to calculating the Universe decay constant $\Gamma$ as we discussed in section 3, much less is known about the mixing angle $\theta$. Because $sin\theta \approx \frac{{c_{12} m_{EW}^2 }}{{c_{22} m_P^2 }}$  the small value of $\theta$ reduces to why is $m_{EW}/m_P$ so small which is the electroweak/gravitation hierarchy problem. Popular approaches to this problem are warped extra dimensions \cite{Randall:2005xy} and M-theory compactifications with $G_2$ holonomy \cite{Acharya:2006ia}. In both approaches there exist two sectors of the theory, the $TeV$ brane and the Planck brane in the warped extra dimension case, or the unbroken supersymmetry sector and hidden supersymmetry breaking sector in the M-theory compactification case. If one identifies the wave functions of these two sectors with the states $\psi_1$ and $\psi_2$ then it is reasonable to connect the mixing parameter to the small hierarchy ratio in some manner.

In the brane-world approach to the hierarchy problem the Friedmann equation is modified to \cite{Maartens:1999hf}\cite{Shiromizu:1999wj}:
\begin{equation}
3m_P^2 \frac{{\dot a^2 }}{{a^2 }} = \lambda  + \rho  + \rho ^2 \frac{1}{{2\tau }} + \frac{{N_{dark - rad} }}{{a^4 }}
\end{equation}
where $\tau$ is the brane tension, $\rho$ is the energy density and $N_{dark-rad}/a^4$ is the dark radiation energy density. The five dimensional bulk cosmological constant $\lambda_{5d}$ is related to the four dimensional cosmological constant $\lambda$ through the relation \cite{Maartens:1999hf}:
\begin{equation}
\frac{\lambda }{{m_P^2 }} = \frac{{4\pi }}{{m_5^3 }}(\lambda _{5d}  + \frac{{4\pi \tau ^2 }}{{3m_5^3 }})
\end{equation}
Where $m_5$ is the five dimensional $TeV$ mass scale which is applied to the gauge hierarchy problem. This five dimensional scale is related to the brane tension through the formula \cite{Maartens:1999hf}:
\[
m_5  = \tau ^{1/6} (\sqrt {8\pi } m_P )^{1/3}
\]
If one imposes the cosmological seesaw relation using this scale $m_5$ one has :
\[
\lambda  = c\frac{{m_5^8 }}{{m_P^4 }}
\]
Finally solving (5.2) for the bulk 5d cosmological constant and using the above to write the $\tau$ and $\lambda$ dependence on $m_5$ only, we have:
\[
\lambda _{5d}  =  - \frac{1}{{144\pi }}\frac{{m_5^9 }}{{m_P^4 }} + \frac{c}{{4\pi }}\frac{{m_5^{11} }}{{m_P^6 }}
\]
The basic point is that in the brane-world scenario the 5d cosmological constant is not of the form $m_5^{10}/m_P^5$ as would have been obtained by imposing the cosmological constant seesaw relation in 5d. Instead one imposes the relation in the 4d sense using the modified Friedmann equation (5.1). Also in the $G_2$ holonomy approach to the hierarchy problem one always has one or more four forms from the dimensional reduction of 11d supergravity whose bosonic sector has a four form. This fits well with the chain inflation scenario discussed in section 3 and 4. Again in this approach one imposes the cosmological seesaw relation in 4d. In 11d the four form is dynamical so that the analysis of the flux dependence on the WDW wave function is much more complicated.

One can also turn the problem around and say that the parameter $sin\theta$ is the overlap between two states $\left\langle {\psi_I } \right|{\psi_2 ,a^3 } \rangle$
 from formula (2.3) and these overlaps are expected to be very small for states with different topology or field content. In $2+1$ dimensional gravity overlaps can be calculated from the initial and final topological data and are exponentially small \cite{Witten:1989sx}\cite{Carlip:1989qq}\cite{Carlip:1994tt}. If the state $\psi_1$ is supersymmetric and $\psi_2$ is nonsupersymmetric then the only way the $\psi_1$ state involves nonsupersymmetric effects is through the mixing. Then the mass hierarchy is understood as arising from the small mixing angle rather than the other way around. String/M-theory seems to allow the possibility of such overlaps or mixing between theories of different types. There are two basic approaches to mixing in string/M-theory (1) cosmological unification of string theory \cite{Hellerman:2006hf} and (2) large-N dual cosmology \cite{Bak:2006nh}. In the first approach a transition between the $0A$ and $bosonic$ string has been recently demonstrated through the use of tachyonic $T$ field background which depends on a lightcone coordinate\cite{Hellerman:2006hf}. In the second approach a transition between the $0A$ 2d string and the $0B$ 2d string was accomplished through an intermediate 3d $M$-theory state using a time dependent Fermi surface formulation of the dual theory \cite{Horava:2005tt}\cite{McGuigan:2006iq}.

In both approaches to transitions in string theory time dependence plays a key role. To introduce time dependence first consider the static $AdS^5$ metric written in the form:
 \begin{equation}
ds^2  = \frac{1}{{w^2 }}( - dt^2  + dx^2  + dy^2  + dz^2  + dw^2 )
\end{equation}
In \cite{Sin:2006pv} a deformation of this metric was considered by introducing functions $N, a, \beta_i$ and modifying (5.3) to be:
\[
ds^2  = \frac{1}{{w^2 }}( - N^2 (w,t)dt^2  + a^2 (w,t)(e^{2\beta _1 (w,t)} dx^2  + e^{2\beta _2 (w,t)} dy^2  + e^{2\beta _3 (w,t)} dz^2 ) + dw^2 )
\]
This form is very similar to the initial cosmological ansatz we used in section 2 except for the dependence of the fifth coordinate $w$. Because of the warping factors the variables $a,\beta_i$ don't obey the usual Einstein equations and instead one has a modified Kasner evolution following  from the 5d gravity equations \cite{Sin:2006pv}. The Kasner evolution in the gravity description is related to elliptic flow in the dual gauge theory \cite{Sin:2006pv}. The basic advantage of the dual gauge description is that if the Kasner evolution hits a singularity the dual gauge description could still be valid \cite{Martinec:2006ak}\cite{She:2005mt}. Also if one twists some of the tori one has a dual gauge description with broken supersymmetry. This is mathematically similar to treating the dual gauge theory at finite temperature. The time dependence means that the (in) and (out) values of the cosmological constant could be different. The dual gauge description can in principle yield a nonperturbative description of cosmological transitions and the mixing parameter in string theory.

\section{Conclusions}
In this paper we have extended the discussion of the cosmological constant seesaw to include the large eigenvalue. We find that the large eigenvalue gives rise to a period of inflation terminated by Universe decay and involves a decay parameter $\Gamma$ and a mixing parameter $\theta$. We examined the cosmological constant seesaw in the context of several models of current interest including chain inflation, inflatonless inflation, string theory  and Universe entanglement. While the decay parameter can be calculated in several cosmological models the mixing parameter is more subtle and is related the electroweak/gravity mass hierarchy problem. It may be possible to turn the problem around and argue that the mass hierarchy is so small because vacuum mixing is inherently a classically forbidden process with very small probability. Although semiclassical WDW methods used in this paper are limited, large N methods may be able to quantitatively describe such transitions nonperturbatively, perhaps even in regions where the  spacetime becomes singular. Finally as the large eigenvalue leads to a period of inflation 
it would be interesting to examine the pattern of density perturbations induced by the model. One may suspect that the pattern would be similar to  that from inflatonless inflation  or chain inflation, however as we saw in examining the probability of transition to the accelerating state, there may be large third quantized corrections in the early Universe.

\section*{Acknowledgments}
I wish to thank Chiara Nappi and Tom Banks  for useful discussions. I would also like to thank Aaron Bergman, Jim Bogan, Lubos Motl and Sean Carroll for valuable suggestions and/or blog postings related to this topic.

\end{document}